\documentclass[]{agujournal2018}
\usepackage{apacite}
\usepackage{url} 
\usepackage{lineno}
\usepackage{amsmath}
\journalname{JGR: Space Physics}

\begin{document}

%
%


\title{MMS Observations of Whistler and Lower Hybrid Drift (LHD) Waves Associated with Magnetic Reconnection in the Turbulent Magnetosheath}

%
%




\authors{Zolt\'an V\"{o}r\"{o}s\affil{1,2}, Emilyia Yordanova\affil{3}, Daniel B. Graham\affil{3}, Yuri V. Khotyaintsev\affil{3}, Yasuhito Narita\affil{1} }

\affiliation{1}{Space Research Institute, Austrian Academy of Sciences, Graz, Austria}
\affiliation{2}{Geodetic and Geophysical Institute, RCAES, Sopron, Hungary}
\affiliation{3}{Swedish Institute of Space Physics, Uppsala, Sweden}





\correspondingauthor{Zolt\'an V\"or\"os}{zoltan.voeroes@oeaw.ac.at}




\begin{keypoints}
\item Magnetic reconnection
\item Turbulence
\item Waves
\end{keypoints}

%
%


\begin{abstract}
Magnetic reconnection (MR) and the associated concurrently occurring waves have been extensively studied at large-scale plasma boundaries, in quasi-symmetric and asymmetric configurations in the terrestrial magnetotail and at the magnetopause. Recent high-resolution observations by MMS (Magnetospheric Multi-Scale) spacecraft indicate that MR can occur also in the magnetosheath where the conditions are highly turbulent when the upstream shock geometry is quasi-parallel. The strong turbulent motions make the boundary conditions for evolving MR complicated. In this paper it is demonstrated that the wave observations in localized regions of MR can serve as an additional diagnostic tool reinforcing our capacity for identifying MR events in turbulent plasmas. It is shown that in a close resemblance with MR at large-scale boundaries, turbulent reconnection associated whistler waves occur at separatrix/outflow regions and at the outer boundary of the electron diffusion region, while lower hybrid drift (LHD) waves are associated with density gradients during the crossing of the current sheet. The lower hybrid drift instability can make the density inhomogeneities rippled. The identification of MR associated waves in the magnetosheath represents also an important milestone for developing a better understanding of energy redistribution and dissipation in turbulent plasmas.

\end{abstract}

%
%

 \section{Introduction}
Collisionless magnetic reconnection (MR) has been most intensively studied through in-situ single- and multi-point measurements at the large-scale boundaries of the Earth's magnetosphere, such as the magnetopause \citep[e.g.][]{burch16} and the magnetotail current sheets \citep[e.g.][]{torbert18}. Plasma waves, over the frequency ranges from ion gyrofrequency to electron plasma frequency, often occur in different regions of MR \citep[e.g.][]{vaivads06}. The waves/turbulence can originate from remote physical processes, but can also be generated by MR during the transformation of magnetic energy to thermal and kinetic energy of particle populations.

The impact of turbulence on MR can be studied in the turbulent magnetosheath downstream of a quasi-parallel shock.
Although the occurrence rate of potentially reconnecting current sheets is high \citep{voros16},
the number of observed MR events or MR signatures is rather limited in the turbulent magnetosheath. Cluster and Themis magnetosheath observations of MR were limited by the time resolution and observed the convection outflow velocity \citep{retino07} or ion outflows only \citep{phan07, oieroset17}.
Turbulent fluctuations reaching the ion/electron scales can introduce constraints limiting the full development of MR structures. The possibility to investigate the tiny structures associated with MR in turbulent plasmas was for the first time enabled by the high resolution measurements of the MMS (Magnetospheric Multi-scale) mission \citep{burch16}.
In some cases MMS observed reconnection signatures in the turbulent magnetosheath, such as thin current sheets, nonzero normal magnetic field (indicating the connection of the oppositely oriented fields across the current sheet), Hall magnetic fields and currents, particle acceleration, etc., however, not accompanied by any reconnection outflows \citep{eriksson18}. In a few cases MR generated electron outflows were observed only \citep{yordanova16, wilder17, phan18}. In some other cases both ion and electron outflows were also observed \citep{voros17, eastwood18}.
Recently, \citet{phan18} argued that the electron only outflows or electron-scale reconnection could occur due to the stochastic motions in a turbulent environment provided that the spatial dimensions of current sheets are longer than the electron diffusion region (EDR), but shorter than the ion diffusion region (IDR). In such a situation the magnetized electrons could form an electron jet, however, the unmagnetized ions might not form an ion jet.
It can be interpreted as an environmental effect when the strong turbulent motions prevent the development of MR related length scales over which the ion flows  develop. Another explanation for the unobserved but eventually existing ion outflows could probably be that the spatial domain of validity of the reconnection coordinate system, in which an ion outflow would be recognizable from the data during a crossing, is limited. In this case turbulent motions would introduce ion flow directional uncertainties only. In any case, the observation of reconnection signatures can be more challenging in turbulence than at large-scale boundaries.

We expect that the observations of the local occurrences of waves could reinforce the identification of MR events in turbulent plasmas. Here, the focus is on the whistler and lower hybrid drift (LHD) waves. At large-scale boundaries (magnetopause or magnetotail current sheets) whistlers were observed in the MR outflow region \citep{khotyain16} and in the separatrix region \citep{lecontel16b, graham16}.
In a reconnection event study \citet{huang16} observed whistler waves simultaneously propagating away from and towards the X-line. The former were generated by temperature anisotropy in the pileup region embedded into reconnection outflow, the letter were observed in the reconnection separatrix region. In a recent statistical study based on Cluster data in the magnetotail \citet{huang17} found that whistler waves are abundant in reconnection separatrix and in flow pileup regions while rare near the X-line. However, Cluster time resolution did not allow to observe electron scales or EDR associated waves.
Large-amplitude whistler waves  were observed by MMS inside the EDR propagating away from the X line \citep{cao17}.
The LHD waves are frequently observed at plasma boundaries with density or/and temperature gradients. LHD waves are often considered in a "2D slab" geometry when the density gradient and the magnetic field are perpendicular and the drift waves propagate approximately perpendicular to both the density gradient and B field \citep{huba78, norgren12, graham17}. Reconnecting current sheets can form 2D slab geometries and plasma configurations sustaining density gradients at their boundaries.

Although whistler waves (or lion roars) are commonly observed in the magnetosheath \citep{baum99} and are not necessarily associated with magnetic reconnection, their observations at separatrices, flow pileup or EDR regions can reinforce the available methodology for identifying MR events in turbulence. For example, this can happen when the turbulent fluctuations make  a straightforward identification of the local coordinate system, in which reconnection physics can be understood, difficult. However, wave observations alone cannot provide evidence for MR, observations of some other signatures, for example reconnnection inflows/outflows, separatrix regions, reconnection electric fields, etc. are needed as well.  It is also important to show that MR in a turbulent environment can be associated with instabilities, spatial gradients and waves similarly to the cases occurring at large-scale boundaries, thoroughly studied mostly in laminar MR cases.

In this paper we investigate the waves associated with a  MR event in the quasi-parallel magnetosheath. The event analyzed here has already been studied thoroughly and  signatures of turbulent MR have been identified \citep{voros17}. After describing the instrumentation (Section 2), the MR event is overviewed and the expected wave activity locations are described (Section 3). Before the summary and conclusions (Section 6) the whistler (Section 4) and LHD waves (Section 5) are analyzed in more detail.
\section{Instrumentation}
The ion and electron moments with time resolution of 150 ms and 30 ms, respectively, are available from Fast Plasma Investigation (FPI) instrument \citep{pollock16}. The electric field data from Electric Double Probes (EDP) instrument are available with time resolution of 8 kHz \citep{ergun16, lindkvist16, torbert16}.  The merged digital fluxgate (FGM) \citep{russell16} and search coil (SCM) \citep{lecontel16a} data were developed by
using instrument frequency and timing models that were created during the FIELDS integration test campaign \citep{torbert16, fischer16}. The merged magnetic data analyzed here consists of FGM measurements
below 4 Hz and data from SCM between 1 Hz and 6 kHz.

\section{MR and wave observations}
On November 30, 2015 MMS observed fluid and kinetic scale signatures of ongoing MR in the turbulent magnetosheath.
Although the whole event described in \citet{voros17}  is not consider here, the main findings about the MR geometry are shortly outlined below.

Figure 1a shows the cartoon of MR crossing in the local current sheet LMN system. The trajectory of the spacecraft is indicated by green dashed line, and the abbreviations EDR and IDR correspond to electron and ion diffusion regions, respectively. Inside the IDR the electron outflows ($V_L^e$) are developed only, while outside the IDR both electron and ion outflows are expected ($V_L^{ie}$). The LMN coordinates are shown on top right and the relative distances between the spacecraft in L-N plane and along M on top left (the reference spacecraft is MMS3 at {\bf{0}} position). Here {\bf{L}} is the main magnetic field direction, {\bf{N}} is the normal to the current sheet and {\bf{M}} is the out-of-plane direction. The LMN coordinates are found on the basis of minimum variance analysis yielding {\bf{L}}=[$-$0.07 $-$0.55 0.83], {\bf{M}} = [0.15 0.82 0.55], and {\bf{N}} = [$-$0.99 0.17 0.03] in geocentric solar ecliptic coordinates (GSE). {\bf{N}} is pointing approximately to -X GSE direction. The minimum variance analysis was performed over the time interval between 00:23:55.97 and 00:23:56.39 UT. Based on multi-point timing the motion of the whole structure in normal direction was obtained, giving $V_N$ = 110 km/s.

Figures 1b-d show the LMN magnetic field components for MMS1-4 spacecraft. The $B_L$ component is positive before 00:23:56 UT, and negative afterwards. There is a -15 nT guide field (see Figure 7 in \citet{voros17}), the fluctuations of $B_M$ around this value represent the Hall magnetic field. $B_N$  changes from predominantly positive to negative across the event.

The perpendicular to magnetic field $V_{i\perp L}$ and $V_{e\perp L}$ compared to ($\mathbf{E}\times \mathbf{B}/B^2)_L$ for MMS3 are shown in Figure 1e. The other spacecraft, when reaching the same locations, show similar flow patterns.  Figure 1f shows the $V_{i\perp N}$ velocities for MMS1-4 spacecraft, the dashed horizontal lines indicate the uncertainty of $\pm$10 km/s with max($V_{i\perp N}$)$ \sim V_N$.

Comparing the magnetic and velocity data (Figures 1b-f) to the cartoon (Figure 1a) it is seen that, initially, between 00:23:52 and 00:23:53.4 UT,  the spacecraft are in the outflow/inflow/separatrix regions. Across of these locations $V_{i\perp L}$, $V_{e\perp L}$ and $(\mathbf{E}\times \mathbf{B}/B^2)_L$ are predominantly positive. $V_{i\perp N}$ inflow velocity, relative to the $V_N$ velocity of the reconnection structure (dashed lines), is
$V^{'}_{N} \sim $ -50 km/s. Between 00:23:53.4 and 00:23:53.8 UT $V^{'}_{N} \sim $ 0 km/s at MMS3 (green line in Figure 1f) and at the same time
$V_{i\perp L} \sim V_{e\perp L} \sim (\mathbf{E}\times \mathbf{B}/B^2)_L \sim$ 50  km/s (Figure 1e) indicating that the spacecraft crossed the +L directional sub-Alfv\`{e}nic outflow, just outside of the IDR in Figure 1a. The local Alfv\`{e}n velocity is about 70 km/s. Later on, when $B_L < \pm$ 50 nT, $V_{i\perp L} \neq (\mathbf{E}\times \mathbf{B}/B^2)_L$ and the ions are demagnetized. Before 00:23:54 UT MMS3 observes $B_L \sim 0$ nT, $B_N$ is reaching 20 nT, $V_{e\perp L}\sim$ 150 km/s and the electrons are demagnetized for a short interval. All this indicates that MMS3 can be close to the EDR. In fact, during this short interval, MMS3 observes also parallel electric fields, $\mathbf{J}.\mathbf{E'} > 10$ $nW/m^3$ and $\sqrt{Q} \sim$ 0.04. Here $\mathbf{J}$ is the current density, $\mathbf{E'}$ is the electric field in the moving frame of electrons and $\sqrt{Q}$ is the dimensionless agyrotropy calculated from the electron pressure tensor. These values might correspond to the outer EDR \citep{voros17}.
MMS3 also observed two electron outflows at 00:23:55.08 and at 00:23:56 UT with $V_{e\perp L} \sim $ -75 km/s, when the spacecraft was inside the IDR (Figures 1a,e).

In Figures 1g-k MMS1-4 observations of the electron density ($N_e$), ion and electron temperatures ($T_i$, $T_e$), the magnitude of current density ($\mathbf{J_P}=N_eq(\mathbf{Vi}-\mathbf{Ve})$; $q$-charge), the magnitude of diamagnetic current ($\mathbf{J_D}=N_eq(\mathbf{V}_{D_i}-\mathbf{V}_{D_e}) = k_B(T_i+T_e)\mathbf{B}\times \mathbf{\nabla} N_e/B^2$; the density gradient is calculated from four-point measurements, $\mathbf{V}_{D_{i,e}}$ - ion, electron diamagnetic drift velocities; $k_B$ - Boltzmann constant) and plasma $\beta$ are shown. The largest density gradients in Figure 1g occur at 00:23:52.5, 00:23:53.9,  between 00:23:54.7 -- 00:23:55.4 and between 00:23:56 -- 00:23:56.7 UT. The density gradients are correlating well with the enhancements of $|\mathbf{J_D}|$.
Since $T_i \gg T_e$ (Figure 1h) the diamagnetic current is mainly determined by the ion diamagnetic drift $\mathbf{V}_{D_i}$.
During the considered time interval the local $\beta$ enhancements (Figure 1k) are correlated with enhanced $\mathbf{J_P}$, $\mathbf{J_D}$ currents (Figures 1i,j) and small values of $B_L$ (Figure 1b), indicating that the large plasma $\beta$ values occur because the spacecraft are getting closer to the current sheet.

Let us first demonstrate that during this event there occur different type of waves over the frequency range of 10 Hz - 4 kHz.
In what follows, the electric field is transformed to the moving frame of the plasma through $\mathbf{E}+\mathbf{V}_p\times\mathbf{B}$, where
$\mathbf{V}_p=[\langle V_{iL}\rangle \;\; \langle V_{iM}\rangle \;\; V_{N}] = [-27 \;\; -143 \;\;110]$ km/s, the L and M components are averages calculated from the ion velocity before and after the event, the N component is obtained from multi-point timing. We note, that the final results on wave properties do not change when instead of $\mathbf{V}_p$ the deHoffmann-Teller velocity is used.
For MMS3 spacecraft, the magnetic field magnitude, the $B_L$, $B_M$ and $B_N$, components are in Figure 2a.
Figures 2b and d show the high-pass filtered (f $>$ 10 Hz) parallel and two perpendicular components (in field aligned coordinates - FAC) of magnetic and electric field fluctuations, respectively. The magnetic and electric dynamic spectra are shown in Figures 2c and e. The polarization properties of the waves including the wavevector in the time-frequency plane are calculated on the basis of singular value decomposition \citep{santolik03}. The Ellipticity of the waves is shown in Figure 2f and the angle between the wavevector and magnetic field ($\theta_k$) in Figure 2g. In the dynamic spectra the overplotted lines correspond to the  proton plasma frequency (red lines): $f_{pp}=(1/2\pi)\sqrt{N_pq^2/m_p\varepsilon_0}$ ($m_p$ - ion mass, $\varepsilon_0$ - permittivity of free space); to the electron cyclotron frequency and its 0.5 and 0.1 times fractions (white dashed lines): $f_{ce}=(1/2\pi)qB/m_e$ ($m_e$ - electron mass); to the lower-hybrid frequency (black lines): $f_{LH}\approx \sqrt{f_{ce}f_{cp}}$ ($f_{cp}$ is the proton cyclotron frequency).

There are broad-band electrostatic fluctuations over the frequency range 100 Hz - 4 kHz ($f \in [0.1f_{ce}\;\; 2f_{pp}]$), Figure 2e) which are absent in the magnetic dynamic spectra (Figure 2c). Narrow band electromagnetic fluctuations are seen over the frequency range of $f \in [0.1f_{ce}\;\; 0.5f_{ce}]$ between 00:23:52.1 and 00:23:52.2 UT and between 00:23:52.4 and 00:23:53.1 UT in both magnetic and electric spectra (Figures 2c, e). For the same time and frequency intervals, the Ellipticity $\sim$ +1 (Figure 2f) and $\theta_k$ is predominantly between 0 and 30 degrees. These electromagnetic fluctuations are right-hand polarized whistler waves propagating along the or under small angle to the magnetic field. Between 00:23:53.1 and 00:23:54.1 UT at around $f = 0.1f_{ce}$ there is electromagnetic power with the same Ellipticity and $\theta_k$, therefore these fluctuations can also be whistler waves.

From 00:23:55 until 00:23:56.6 UT the largest electric field power is around or below the lower-hybrid frequency range (black line in Figure 2e) and over the same frequencies the Ellipticity is predominantly between -1 and 0, occassionally between 0 and +1. $\theta_k$ is nearly 90 degrees around $f=f_{LH}$, indicating perpendicular to magnetic field wave vectors. These fluctuations can correspond to LHD waves. At around 00:23:55.8 and 00:23:56.4 UT there are short, $\sim$ 0.2 s intervals, where at $f \sim 0.1f_{ce}$, the Ellipticity is +1 and $\theta_k \sim 0$ (Figures 2 f,g). These fluctuations can again correspond to whistler waves.

\section{Whistler waves}
The event before 00:23:55 UT in Figure 2, where predominantly the whistler emissions occur, is shown in Figure 3. The magnetic field magnitude, the LMN magnetic components (Figure 3a), the magnetic and electric field fluctuations in FAC coordinates (Figures 3b, d) and the corresponding dynamic spectra (Figures 3c, e) are shown in more detail. The parallel and perpendicular to magnetic field electron temperatures, $Te_{||}$ (black line) and $Te_{\perp}$ (red line) are shown in Figure 3f. Between 00:23:52 and 00:23:52.7 UT the occurrence of whistler waves can be understood on the basis of temperature anisotropy, when  $Te_{perp} > Te_{||}$. However, after 00:23:52.7 UT, until 00:23:55 UT $Te_{\perp} \leq Te_{||}$. Figure 3h  shows the high (250 - 1000 eV) and Figure 3i the low energy (70-250 eV) electron pitch angle distribution spectrograms. Both low and high energy electrons show perpendicular populations at the beginning of the interval when also $Te_{\perp} > Te_{||}$. Between 00:23:52.7 and 00:23:54.2 UT the low energy electron population, which has the largest contribution to $Te$, becomes   parallel and antiparallel to the magnetic field (Figure 3i). Field aligned electrons are energized by parallel electric field at the boundary of the EDR at around 00:23:54 UT (see \citet{voros17}). The resonant energies for the whistler waves can be estimated from the resonance condition $V_{\parallel res}=(f-f_{ce})\lambda_{\parallel}$, where $\lambda_{\parallel}=V^w_{ph}/f$ is the parallel wavelength. Between 00:23:52 and 00:23:53 UT the whistler phase velocity $V^w_{ph}$ was estimated from the ratio $|\mathbf{E}/\mathbf{B}|$ giving $V^w_{ph}=$1600 $-$ 3200 km/s. For these high velocities the Doppler effect can be neglected. $\lambda_{\parallel}$ was between 6 and 16 km and the resonant energies between 150 and 1400 eV. Between 00:23:53.2 and 00:23:54.4 UT the $|\mathbf{E}/\mathbf{B}|$ velocity was strongly fluctuating between 1000 and 10000 km/s, leading to $\lambda_{\parallel}= 10 - 100$ km and energies between 200 eV and 14 keV. The energy range of 250 - 1000 eV in the pitch angle distribution in Figure 3h roughly covers the lower end of the resonant energies, though much larger than the average electron temperature which is about 45 eV (Figure 1h). Therefore, when $Te_{\perp} \leq Te_{||}$, the observed whistler emissions can be associated with the perpendicular population of high energy electrons.

Finally, Figures 3g and j show the parallel normalized Poynting fluxes ($S_{||}/|S|$) for MMS3 and MMS1 spacecraft, respectively. Positive values of $S_{||}/|S|$  correspond to waves propagating parallel to the magnetic field, while negative values indicate antiparallel propagation.

MMS3 (Figure 3g) observed waves over 0.1-0.5 $f_{ce}$ propagating along the magnetic field lines ($S_{||}/|S|\sim 1$)
between 00:23:52.4 and 00:23:53.1 UT. At around 00:23:53.25 and 00:23:53.85 UT there are two short ($\sim$ 0.1 s long) time intervals where $S_{||}/|S|\sim$0 and the propagation direction cannot be determined. At the same times broad-band electrostatic noise is visible (Figure 3e), which locally can grow faster than whistler mode waves heating the parallel electron populations \citep{zhang99}. In fact, both time intervals are associated with $Te_{||} > Te_{\perp}$ (Figures 3f, i).

Now we consider whistler waves propagating in the antiparallel to magnetic field direction. In Figure 3g, between 00:23:52 and 00:23:52.35 UT, over the frequency range of 0.1-0.5$f_{ce}$, $S_{||}/|S|\sim -1$. Afterwards, until 00:23:52.9 UT, there is an overlapping time interval when also $S_{||}/|S|\pm$ 1. The +1 values occur closer to 0.5$f_{ce}$ and the -1 values are closer to 0.1$f_{ce}$ (dashed white lines). These time intervals correspond to inflow/separatrix regions in Figures 1a-f.
MMS1 is almost at the same L and M coordinates as MMS3, the largest separation ($\sim $21 km) is in N direction (Figure 1a). Therefore, MMS1 is more towards the inflow region than MMS3 and during the considered time interval observes whistler waves in the antiparallel direction only ($S_{||}/|S|$ -1, Figure 3j). These whistler waves can propagate in the antiparallel direction from a remote source and at MMS3 positions can mix with the whistlers propagating in parallel direction from the outflow or EDR source regions (Figures 3g, j).

After 00:23:53.2 UT at MMS1 (Figure 3j) and 00:23:53.45 UT at MMS3 (Figure 3g) an alternating pattern of $S_{||}/|S|=\pm 1$ is seen near 0.1$f_{ce}$ and between 0.1-0.5$f_{ce}$. The whistler waves propagate in parallel and antiparallel directions to the magnetic field. It can be seen in Figure 1e that after 00:23:53.45 UT the MMS3 spacecraft first enters the reconnection outflow, then crosses the electron outflow inside IDR at 00:23:53.9 UT. Between 00:23:54 and 00:23:54.35 UT MMS3 crosses the outer region of EDR. These locations, indicated by red wave double arrows in Figure 1a, can be the source regions of whistlers \citep{zhang99, cao17}. The reason is that whistler waves generated by the temperature anisotropy of electrons can simultaneously propagate into opposite directions within their source region \citep{zhang99}. Noticeably, between 00:23:54.25 and 00:23:54.45 UT, MMS1 observes waves propagating again into antiparallel direction (Figure 3j). Since MMS1 is northward from MMS3, the observed whistlers  propagate in direction antiparallel to the field lines in the inflow region.

In summary, the propagation directions of the whistler waves are indicated in Figure 1a by red wave arrows. The whistler waves propagate in the antiparallel and parallel directions in the inflow/separatrix regions. The whistlers in antiparallel direction propagate towards the X-line from remote sources and are observed more downstream at the border of the $+V_L^{ie}$ outflow and also closer to the X-line in the $-V_N'$ inflow region. Those whistlers propagating along the magnetic field lines, again downstream at the border of the $+V_L^{ie}$ outflow, can be generated in the source region of whistler waves which is closer to the X-line and the magnetic equator. The red double arrow waves in Figure 1a indicate the possible locations of the whistler source regions where the waves propagate in parallel-antiparallel directions closer to the magnetic equator within the outflow and  in the outer EDR regions. We mention that, for simplicity, whistler waves occurring in the $-V_L$ outflow region observed for very short time intervals, are not shown in Figure 1a.


\section{LHD waves}
The event after 00:23:55 UT in Figure 2, where predominantly the LHD waves are expected, is shown in more detail in Figure 4. MMS3 and MMS1 observations of the magnitude of magnetic field and LMN magnetic components (Figures 4 a,d), the LMN electric field fluctuations (Figures 4 b,e) and the scalar potentials calculated from the electric and magnetic fields ($\phi_E$, $\phi_B$, Figures 4 c,f) are compared.

Between 00:23:55.2 and 00:23:56.9 UT the average $f_{LH}\sim$ is 40 Hz and the strongest fluctuations occur over the range of $\sim$ 60 Hz (Figure 2e), therefore a band-pass filter is applied over the frequency range of 10-70 Hz. The phase velocity of LHD waves $\mathbf{V}^{LH}_{ph}$ can be determined by a single spacecraft method using the scalar potentials \citep{norgren12}. The wave potential can be calculated over the LHD frequency range through $\phi_B = |\mathbf{B}|\delta B_{||}/(qN_e\mu_0)$, where $\delta B_{||}$ are the band-pass filtered parallel magnetic fluctuations, $\mu_0$ is the permeability of free space. The potential can also be calculated from $\phi_E=\int \delta \mathbf{E}.\mathbf{V}^{LH}_{ph} dt$ along the trajectory of the spacecraft. The phase velocity is found by least-square fitting of $\phi_E$ and $\phi_B$ and the best wave propagation direction in the plane perpendicular to the magnetic field is found by a correlation method \citep{norgren12}.

During the time period considered in Figure 4, large density gradients are present (Figure 1g) and plasma $\beta$ drops down and fluctuates between 3 and 15 (Figure 1k). The average time delay between the MMS3-MMS1 magnetic field signatures is $\sim$0.3 s (Figures 4a,d) and when the plasma $\beta$ approaches 10 (Figure 1k) roughly at 00:23:55.1 UT (MMS3) and at 00:23:55.4 UT (MMS1), the electric field and scalar potential fluctuations start immediately to develop (Figures 4 b,c and e,f). This indicates that the fluctuations are generated at boundaries with density gradients when $\beta$ becomes smaller. The lower hybrid drift instability (LHDI) generating the LHD waves is getting stabilized when $\beta \geq 1$.
However, for $T_e<T_i$  the stabilization is weaker than for $T_e=T_i$ \citep{davidson77}. In our case $T_i/T_e$ changes from 5 to 10, increasing towards the end of the time interval (Figure 1h). Statistical analysis of MR events in the Earth's magnetotail shows that the magnitude of electric field fluctuations over the LHD frequency range increases when $0.1\leq\beta\leq$10 \citep{zhou14}. In the tail plasma sheet the ratio $T_i/Te$ varies between 2 and 10 \citep{wang12}, which is similar to our case. The free energy supporting LHDI comes from magnetic field and density inhomogeneities and the LHD waves propagate perpendicular to both magnetic field and density gradient directions \citep{krall71}. The controlling parameter is the ratio $L_n/\rho_i$ , where $L_n=(\partial ln N/\partial x)^{-1}$ is the density gradient length scale and $\rho_i$ is the ion gyroradius.
To excite the LHDI sharp density gradients are needed, which leads to the condition $L_n/\rho_i \leq (m_i/m_e)^{1/4}=$ 6.5 \citep{huba78}. The density gradient is also related to the diamagnetic drift through $L_n/\rho_i=V_{thi}/(2V_{Di})$, where $V_{thi}$ is the ion thermal velocity and  $\mathbf{V}_{Di} = \mathbf{V}_{\perp i}-\mathbf{E}\times \mathbf{B}/B^2$ is the ion diamagnetic drift velocity.

Figures 4g-i show MMS1-4 observations of the M and N components of $V_{Di}$,  $V_{thi}/(2V_{Di})$ and $L_n/\rho_i$, respectively.
$V_{thi}/(2V_{Di})\leq$ 0.5 (Figure 4h) slightly enhanced at the current sheet crossing only (Figures 4 a,d). This means that $V_{Di}\geq V_{thi}$ and since the growth rate of LHDI is $\sim (V_{Di}/V_{thi})^2$ \citep{freidberg77}, the instability can be sustained.
Here $L_n$ was estimated through
$\langle N_e\rangle \triangle X_{ij}/\triangle Ne_{ij}$, where $\triangle(X, Ne)_{ij}$ are differences of distances and densities between the spacecraft $i,j$, and $\langle N_e\rangle$ is the average electron density. Figure 4i shows $L_n/\rho_i$ when $L_n$ is calculated between pairs MMS1-3 (green line) and MMS2-4 (blue line), respectively.
MMS1-3 separation is mainly in N, MMS2-4 separation is mainly in M direction (Figure 1a). $L_n/\rho_i$ in M direction (blue line) shows large fluctuations, indicating that there exist density inhomogeneities also in the out-of-plane direction.
$L_n/\rho_i$ in normal direction (green line) fluctuates strongly between 1 and 100, and does not match $V_{thi}/(2V_{Di})$.

The density inhomogeneities (Figure 4i) show fluctuations with a quasi-period of 0.1-0.3 s, indicating that, as the spacecraft cross the structures, the density boundary might be rippled in space.
Recently, a rippled density structure generated by LHDI has also been found at dipolarization fronts in the Earth's magnetotail \citep{pan18}. Using a simple model \citet{pan18} demonstrated that a rippled density boundary leads to a fluctuating $B_L$ magnetic field in N direction resulting in phase velocity of the LHDI waves with significant component in N direction. When the $B_L$ magnetic field components in our event are time shifted to a reference MMS spacecraft, there also exist short duration $\sim 10$ nT fluctuations seen in $B_L$ between the MMS spacecraft (not shown). The rippled density boundaries can explain the observed N directional propagation of the LHDI waves (see below). Since the quasi-period of $\phi_{E,B}$ fluctuations (Figure 4f) is about 0.1 s this might indicate that the density inhomogeneity boundary is rippled due to LHDI, which is also exciting the LHD waves. Nevertheless, in the turbulent magnetosheath, there might exist significant density fluctuations introduced by turbulent fluctuations which are not studied here.
By using the average  $V_p$, 0.1 s correspond to spatial size of 18 km, comparable to the separation of MMS1-3 in N direction. This means that when a rippled density structure is crossed MMS1-3 is expected to see strong fluctuations of $L_n/\rho_i$ corresponding to subsequent local sharp and weak density gradients. When the strongest gradients are considered $L_n/\rho_i$ is between 1-10. Taking also into account the uncertainties in determining $L_n$, the  independently calculated measures in Figures 4 h,i suggest that the density gradients can sustain LHDI \citep{freidberg77}, however, the instability may operate between the strong ($V_{Di} > V_{thi}$) and weak ($V_{Di} < V_{thi}$) drift regimes \citep{norgren12}.

The phase velocity $\mathbf{V}^{LH}_{ph}$ is calculated using the single-point method of scalar potentials \citep{norgren12}, described above. First, for each spacecraft, $\mathbf{V}^{LH}_{ph}$ was determined for the intervals with enhanced $\delta E$ or $\phi_{E,B}$ fluctuations. For example, for MMS1, it is the time interval between 00:23:55 and 00:23:56.7 UT (Figures 4 b,c). The correlation coefficients $C_{\phi}$ were between 0.78 and 0.91. The direction of $\mathbf{V}^{LH}_{ph}$ was also determined as a maximum variance direction (MVD) of $\delta \mathbf{E}$ \citep{pan18}. The ratio of maximum to intermediate eigenvalues, $\lambda_{max}/\lambda_{int}$ was between 2.5 and 4. The results are in Table 1 where also differences in propagation directions obtained from the two methods are shown. At MMS1, 3 the differences are acceptable, 7-12 degrees, at MMS2, 4 are much larger, 23-32 degrees. However, at each spacecraft the propagation is predominantly in +N direction, instead of the M direction expected for the LHD waves in a slab geometry. The phase velocity is changing between 71 and 95 km/s, comparable to $\mathbf{V}_p$. Therefore, the strongest fluctuations in the dynamical spectrum are spread around $f_{LH}$ due to the Doppler shift in Figure 2e.

The estimation of the LHD wave properties for a longer time interval ($\sim$1.5 s) offers an interpretation based on the LMN system in which the MR geometry can be understood. The occurrence of rippled density inhomogeneities due to LHDI can already introduce LHD wave propagation directions not typical for the slab geometry \citep{pan18}. Moreover, as seen from the $\delta E$ and $\phi_{E,B}$ fluctuations (Figures 4 b,c,e,f) the time series are evidently nonstationary. For example, the perpendicular fluctuations $\delta E_{M,N}$ prevail except for  shorter time intervals near the largest Hall magnetic field $B_M$ at 00:23:56.05 UT (MMS3, Figure 4a) and at 00:23:56.25 UT (MMS1, Figure 4d), where also the amplitude of $\delta E_L$ or $\delta E_{L,M}$ fluctuations become larger.  All this indicates that locally the wave vector changes directions which introduces errors to the estimation of $\mathbf{V}^{LH}_{ph}$ direction over a longer time interval. In fact, when LHD wave characteristics are estimated over shorter time intervals, e.g 0.2 s, both $C_{\phi}$ and $\lambda_{max}/\lambda_{int}$ increase and outside of the prevailing Hall magnetic field interval, the wave propagation direction remains +N as in Table 1. There are some variations in L and M directions indicating that the wave vector points into N direction with some degrees of freedom between L and M directions. However, when $B_M$ is reaching $\sim$-50 nT near the current sheet crossing, for a short time, the wave vector direction is turning towards L-M directions. Also, near the separatrix region where $B_L<$-60 nT the lower frequency $\delta E_M$ component dominates (Figures 4 b,e). Here the amplitude of the corresponding potentials is small. From MVD the estimated average direction of the wave vector is +M: [0.1 0.96 0.2], which changes by 4-6 degrees between MMS1-2-3 and by 23 degrees at MMS4. Figure 1a shows  that MMS4 is the outermost spacecraft in -M direction, therefore the observed differences can be due to a spatial effect.

The propagation directions of the LHD waves are indicated in Figure 1a by magenta wave arrows.

\section{Summary and conclusions}
The main goal of this paper was to investigate the association of MR and waves in a turbulent environment. It has been shown that whistler waves seen near the separatrix/inflow region can propagate in anti-field-aligned direction from remote sources and in field-aligned direction propagating out from the source regions which are located closer to the magnetic equator and X-line within the outflow and outer EDR. In the outflow and EDR regions whistlers may be generated by  the temperature anisotropy of the main and high-energy electron populations.

These findings are fully consistent with the results of \citet{huang16} who observed two types of whistler waves  associated with MR in the Earth's magnetotail. The first type of whistlers was generated by electron temperature anisotropy at the magnetic equator within the pileup region of reconnection outflow. These waves propagated downstream along the magnetic field. The second type of whistlers was observed in the reconnection separatrix region propagating in anti-field-aligned direction towards the X-line.
\citet{huang16} speculated that these whistlers might be generated by the electron beam-driven whistler instability or \u{C}erenkov emission from electron phase-space holes.
A recent statistical study based on Cluster data in the magnetotail further confirmed that whistler waves occur mainly in the reconnection separatrix and flow pileup regions, however, whistler occurrence within the EDR could not be straightforwardly studied because of the time resolution of Cluster \citep{huang17}.

Whistler waves were also observed at the magnetopause near or inside the EDR propagating away from the X line \citep{tang13, cao17}. High resolution MMS observations showed that the resonance energy for anisotropic electron population inside EDR was $\sim$300 eV   with maximum growth rate at $f=$0.2$f_{ce}$ \citep{cao17}. In our case, at the outer EDR a parallel electric field is heating mainly the field-aligned low energy electrons ($<$250 eV), leading to $Te_{||}>Te_{\perp}$, however, the perpendicular anisotropy of higher energy electrons ($>$250 eV) may generate the observed whistlers with maximum emissions around $f=$0.1$f_{ce}$ (Figure 3).
Noticeably, the locations and the propagation directions of whistler waves in our turbulent MR event (Figure 1a) are the same as in the above event and statistical magnetotail and magnetopause reconnection studies.

The LHD waves excited by LHDI occur when the spacecraft cross density gradients, the plasma $\beta$ decreases below 10, $T_i/T_e$ increases and the ion diamagnetic drift velocity is comparable to or slightly larger than the ion thermal velocity. This happens in the vicinity of -L directional electron outflows and during crossing of the current sheet. The density gradient length scale normalized to the proton gyroradius indicates rippled density inhomogeneities at MR boundaries. Since the fluctuations of density gradient are accompanied by fluctuations of the magnetic field  and the  scalar potential varies over the same time scales, we speculate that both the waves and the density inhomogeneities may be generated by the same LHDI.
The LHD waves propagate predominantly to N direction during the considered time interval. However, the Hall magnetic field dominating for a short time introduces another deviation from the slab geometry. As a consequence, the wave vector is locally (at large $B_M$) pointing to L-M directions. Finally, at the -$B_L$ separatrix region the wave vector is locally M directional.

In summary, whistler and LHD waves can occur in different regions of MR embedded into turbulent and compressional magnetosheath  resembling the large-scale boundary reconnection cases. In the turbulent MR event studied here the whistlers occurred in the separatrix, reconnection outflow and outer EDR regions. Although whistler observations in the EDR are rare \citep{tang13, cao17}, whistler waves in the separatix and outflow regions are frequently observed at large-scale boundary reconnection case studies \citep{khotyain16, huang16, lecontel16b, graham16}. More importantly, it has been shown in a statistical study \citep{huang17} that whistlers generated mainly by electron temperature anistropy frequently occur near reconnection separatrix and outflow pileup regions in the Earth magnetotail. At large-scale boundary MR events LHD waves propagate in the out-of-plane (M) direction, which is perpendicular to both the density gradients and magnetic field \citep{norgren12, graham17}. These reconnecting current sheets correspond to the '2D slab geometry' configuration. However, when LHD waves are observed at the dipolarization fronts of reconnection outflows in the magnetotail, the density boundaries are rippled, the slab geometry is perturbed and the propagation direction of LHD waves is more along the normal direction to the current sheet \citep{pan18}. In the turbulent MR event we also observed rippled density boundaries which were associated with LHD waves propagating to the normal direction of the current sheet. However, the generation mechanism(s) of boundary perturbations is (are) not clear. It can be the LHD instability  or/and turbulent mixing which is responsible for rippled boundaries. The latter is naturally occurring at turbulent boundary layers.

Both whistler and LHD waves can be generated by instabilities occurring in different locations of MR. In this paper it was shown that the same instabilities which generate waves at large-scale boundary MR events can occur at reconnecting current sheets in turbulence. The waves and instabilities can affect MR in multiple ways, for example, generating anomalous resistivity and accelerating MR. However, it requires further investigations to show that the waves and instabilities could affect MR in turbulent plasmas in the same way as at large-scale boundaries.

 \begin{figure}[h]
 \includegraphics[width=12cm]{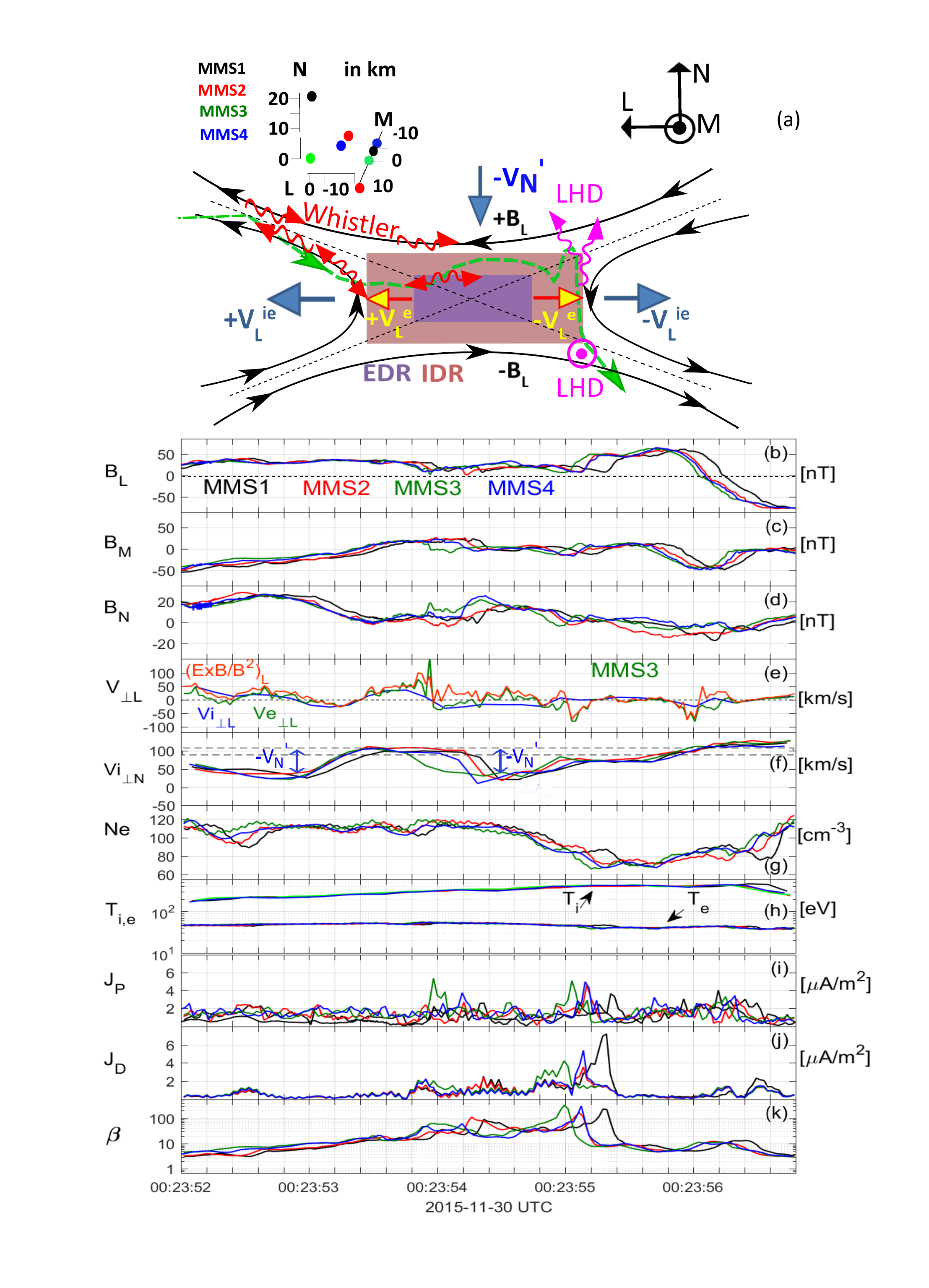}
 \caption{Event overview. (1a) The LMN coordinate system, crossing of the reconnection and the relative positions of spacecraft;
 (1b-k) MMS1-4 data, except for panel e, which contains MMS3  data only;
 (1b-c-d) L,M,N components of the magnetic field
 (1e) L components of perpendicular to magnetic field ion and electron velocities and the L-component of the convection velocity for MMS3;
 (1f) N components of perpendicular to magnetic field ion velocities, -$V'_N$ is the inflow velocity;
 (1g) electron densities;
 (1h) ion and electron temperatures;
 (1i) current density calculated from local plasma parameters;
 (1j) diamagnetic current;
 (1k) plasma $\beta$;
 The propagation directions of whistler and LHD waves are shown by red and magenta wave arrows and +M directional vector, respectively.

 }
 \label{figone}
  \end{figure}

\begin{figure}[h]
 \includegraphics[width=12cm]{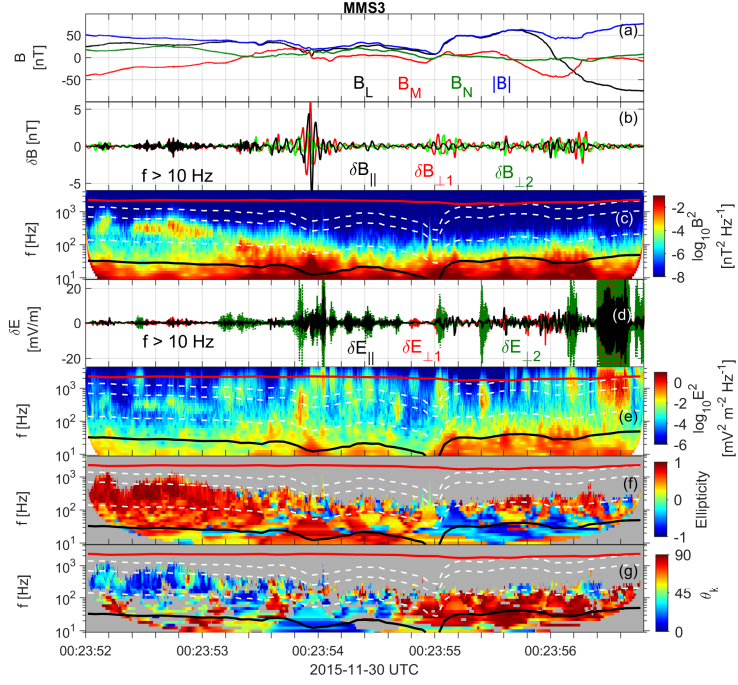}
 \caption{Wave analysis.
 (2a) L,M,N components and the magnitude of the magnetic field;
 (2b) high-pass ($>$10 Hz) filtered magnetic field fluctuations (parallel and perpendicular) in FAC coordinates;
 (2c) dynamic spectrum of $B^2$ fluctuations;
 (2d) high-pass ($>$10 Hz) filtered electric field fluctuations (parallel and perpendicular) in FAC coordinates in the moving frame of plasma;
 (2e) dynamic spectrum of $E^2$ fluctuations;
 (2f) Ellipticity;
 (2g) the angle between the wave vector and the magnetic field.
 The overplotted lines in time-frequency planes correspond to:
$f_{pp}$ - proton plasma frequency (red line); $f_{ce}$ -electron cyclotron frequency and its 0.5 and 0.1 fractions (white dashed lines);
$f_{LH}$ - lower hybrid frequency (black line).
 }
 \label{figtwo}
  \end{figure}

\begin{figure}[h]
 \includegraphics[width=12cm]{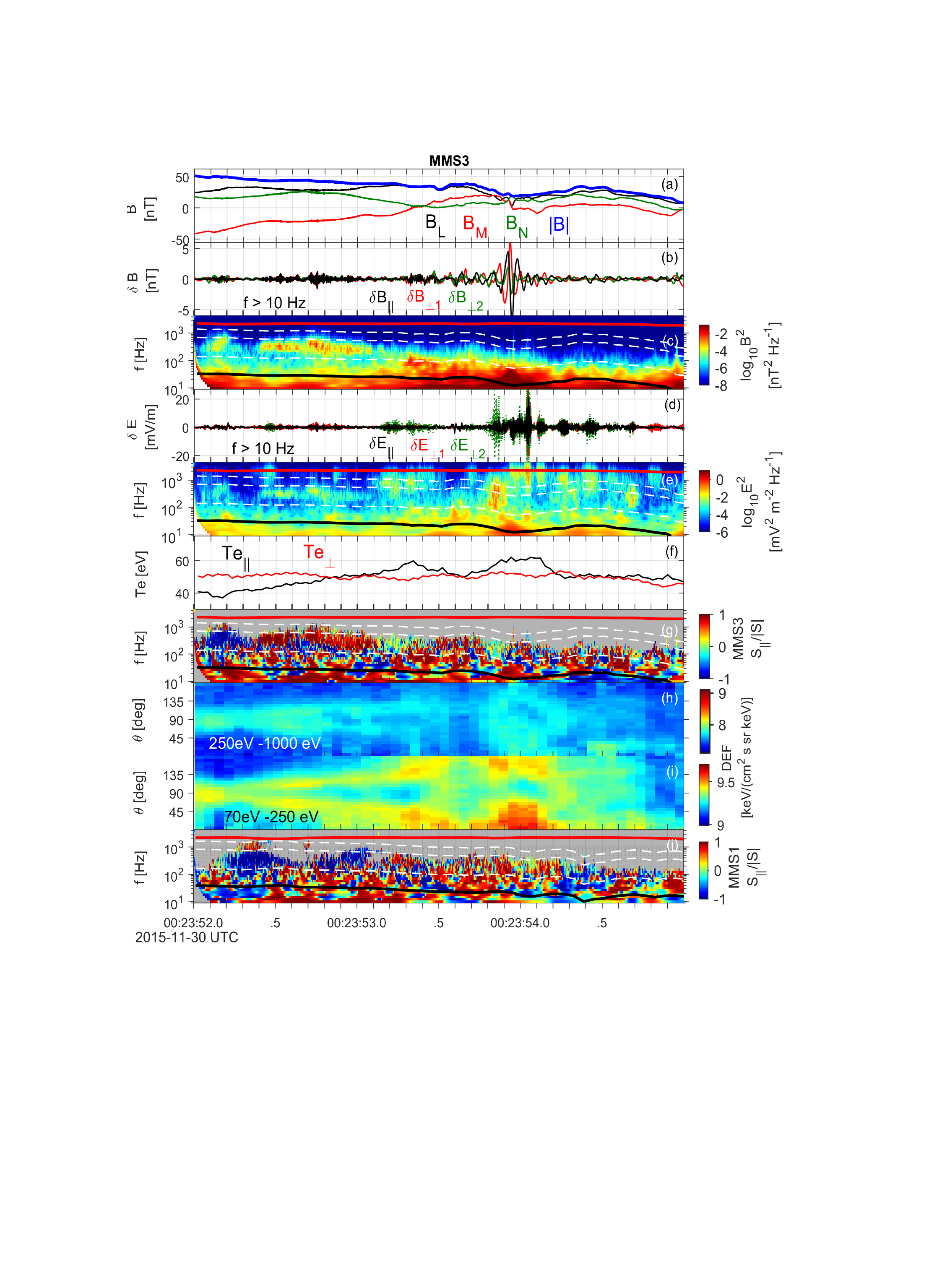}
 \caption{Whistler wave analysis.
 (3a-i) MMS3 data, panel j contains MMS1 data;
 (3a) L,M,N components and the magnitude of the magnetic field;
 (3b, d) high-pass ($>$10 Hz) filtered magnetic and electric field fluctuations (parallel and perpendicular) in FAC coordinates;
 (3c, e) dynamic spectra of $B^2$ and $E^2$ fluctuations;
 (3f) parallel and perpendicular electron temperatures;
 (3g) parallel normalized Poynting flux from MMS3;
 (3h) high-energy pitch-angle distribution
 (3i) low-energy pitch-angle distribution
 (3j) parallel normalized Poynting flux from MMS1;
 The red, white-dashed and black overplotted lines correspond to $f_{pp}$, $f_{ce}$ fractions and $f_{LH}$ local frequencies (same as in Figure 2).
  }
 \label{figthree}
  \end{figure}

\begin{figure}[h]
 \includegraphics[width=12cm]{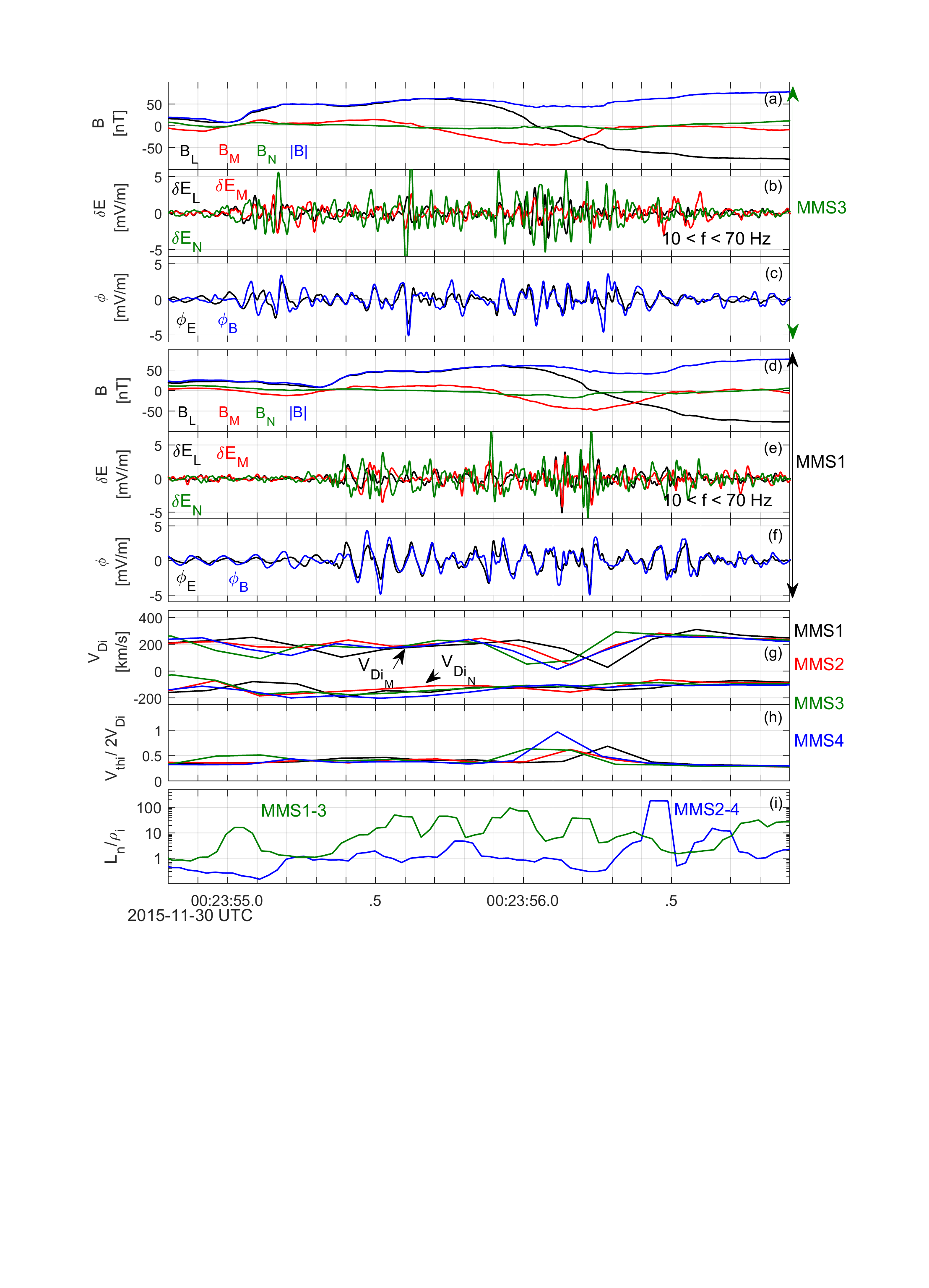}
 \caption{LHD wave analysis.
 (4a-c) MMS3 data;
 (4d-f) MMS1 data;
 (4g-i) MMS1-4 data;
 (4a,d)  L,M,N components and the magnitude of the magnetic field from MMS3 and MMS1 data;
 (4b,e) band-pass (10$<$f$<$ 70 Hz) filtered L,M,N components of electric field fluctuations in the moving frame of plasma;
 (4c,f) scalar potentials calculated from the electric and magnetic fields;
 (4g) M and N components of the ion diamagnetic drift velocity;
 (4h) half of the ratio between ion thermal and ion diamagnetic drift velocities;
 (4i) the normalized density gradient scale.
 }
 \label{figfour}
  \end{figure}

 \begin{table}
 \caption{The phase velocities of LHD waves and their propagation directions estimated by using the method of scalar potentials (abbreviated as $\phi$) \citep{norgren12} and the method of maximum variance direction (MVD) of electric field fluctuations \citep{pan18}.}
 \centering
\begin{tabular}{|l|l|l|l|l|}
  \hline
  MMS & $V^{LH}_{ph}$ (km/s) & LMN direction from $\phi$ & LMN direction from MVD & angle($\phi$-MVD) ($^\circ$) \\ \hline
  1 & 88 & [-0.19 -0.49 0.85]  & [-0.19 -0.38 0.9] & 7\\
  2 & 74 & [0.35 -0.11 0.93] & [-0.04 -0.03 0.99] & 23\\
  3 & 71 & [-0.18 -0.28 0.94] & [-0.22 -0.07 0.97] & 12\\
  4 & 95 & [-0.1 0.13 0.98] & [-0.31 -0.37 0.88] & 32\\
  \hline
\end{tabular}
\end{table}

\acknowledgments
Z.V. and Y.N. were supported by the Austrian FWF under contract P28764-N27 and Austrian FFG ASAP12 SOPHIE, under contract 853994. E. Y. was supported by the Swedish
Civil Contingencies Agency, grant 2016-2102. The data used in this paper are freely available from the MMS data center: https://lasp.colorado.edu/mms/sdc/public/.


%

\begin{thebibliography}{}

\bibitem[\textit{Baumjohann et~al.}(1999)]{baum99}
Baumjohann W., R.~A. Treumann, E. Georgescu, G. Haerendel, K.-H. Fornacon, and U. Auster (1999).
Waveform and packet structure of lion roars.
\textit{Ann.Geophys.}
\textit{17}, 1528–-1534.

\bibitem[{\textit{Burch et~al.}}(2016)]{burch16}
Burch, J.~L., R.~B. Torbert, T.~D. Phan, L.-J. Chen, T.~E. Moore, R.~E. Ergun, J.~P. Eastwood, D.~J. Gershman, P.~A., et~al. (2016).
Electron-scale measurements of magnetic reconnection in space.
\textit{Science}
\textit{352}, af2939, doi: 10.1126/science.aaf2939

\bibitem[\textit{Cao et~al.}(2017)]{cao17}
Cao, D., H.~S. Fu, J.~B. Cao, T.~Y. Wang, D.~B. Graham, Z.~Z. Chen, et~al. (2017).
MMS observations of whistler waves in electron diffusion region.
\textit{Geophys.Res.Lett.}
\textit{44}, 3954–-3962, doi:10.1002/2017GL072703.

\bibitem[\textit{Davidson et~al.}(1977)]{davidson77}
Davidson, R.~C., N.~T. Gladd, C.~S. Wu, and J.D. Huba (1977).
Effects of finite plasma beta on the lower-hybrid-drift instability.
\textit{Phys.Fluids}
\textit{20}, 301--310, doi: 10.1063/1.861867.


\bibitem[\textit{Eastwood et~al.}(2018)]{eastwood18}
Eastwood, J.~P., R. Mistry, T.~D. Phan, S.~J Schwartz, R.~E. Ergun, J.~F. Drake, et~al. (2018).
Guide field reconnection: exhaust structure and heating.
\textit{Geophys.Res.Lett.}
\textit{45}, 4569–-4577, https://doi.org/10.1029/2018GL077670.

\bibitem[\textit{Ergun et~al.}(2016)]{ergun16}
Ergun, R.E., S. Tucker, J. Westfall, K.A. Goodrich, D.M. Malaspina, D. Summers,  et~al. (2016).
The Axial Double Probe and Fields Signal Processing for the MMS Mission.
\textit{Space Sci. Rev.}
\textit{199}, 167--188, doi:10.1007/s11214-014-0115-x.

\bibitem[\textit{Eriksson et~al.}(2016a)]{eriksson18}
Eriksson, E., A. Vaivads, D.B. Graham, A. Divin, Yu.~V. Khotyaintsev, E. Yordanova, et~al. (2018).
Electron energization at a reconnecting magnetosheath current sheet.
\textit{Geophys.Res.Lett.}
\textit{45}, 8081–-8090, https://doi.org/10.1029/2018GL078660.

\bibitem[\textit{Fischer et~al.}(2016)]{fischer16}
Fischer, D., W. Magnes, C. Hagen, I. Dors, M.W. Chutter, J. Needell, et~al. (2016).
Optimized merging of search coil and fluxgate data for MMS.
\textit{Geosci. Instrum. Method. Data Syst.}
\textit{5}, 521--530, doi:10.5194/gi-5-521-2016.

\bibitem[\textit{Freidberg and Gerwin}(1977)]{freidberg77}
Freidberg, J.~P., and R.~A. Gerwin (1977).
Lower hybrid drift instability at low drift frequences.
\textit{Phys. Fluids}
\textit{20}, 1311--1315, doi: 10.1063/1.862013.


\bibitem[\textit{Graham et~al.}(2016)]{graham16}
Graham, D.B., A. Vaivads, Yu.~V. Khotyaintsev, and M. Andre\'{e} (2016).
Whistler emission in the separatrix regions of assymetric magnetic reconnection.
\textit{J.Geophys.Res.}
\textit{121}, 1934–-1954, doi:10.1002/2015JA021239.

\bibitem[\textit{Graham et~al.}(2017)]{graham17}
Graham, D.B., Yu.~V. Khotyaintsev, C. Norgren, A. Vaivads, M. Andre\'{e}, S. Toledo-Redondo, et al. (2016).
Lower hybrid waves in the ion diffusion region and magnetospheric inflow regions.
\textit{J.Geophys.Res.}
\textit{122}, 517–-533, doi:10.1002/2016JA023572.


\bibitem[\textit{Huang et~al.}(2016)]{huang16}
Huang, S.~Y., H.~S. Fu, Z.~G. Yuan, A. Vaivads, Y.~V.Khotyaintsev, A.Retino, et~al. (2016).
Two types of whistler waves in the hall reconnection region.
\textit{J.Geophys.Res.}
\textit{121}, 6639–-6646, doi:10.1002/2016JA022650.

\bibitem[\textit{Huang et~al.}(2017)]{huang17}
Huang, S.~Y., Z.~G. Yuan, F. Sahraoui, H.~S Fu, Y. Pang, M. Zhou, et~al. (2017).
Occurrence rate of whistler waves in the magnetotail reconnection region.
\textit{J.Geophys.Res.}
\textit{122}, 7188–-7196, doi:10.1002/2016JA023670.

\bibitem[\textit{Huba et~al.}(1978)]{huba78}
Huba, J.~D., N.~T. Gladd, and K. Papadopoulos, (1978).
Lower-hybrid-drift turbulence in the distant magnetotail.
\textit{J.Geophys.Res.}
\textit{83}, A11, 5217–-5226.

\bibitem[\textit{Khotyaintsev et~al.}(2016)]{khotyain16}
Khotyaintsev, Yu.~V., D.B. Graham, C. Norgren, E. Eriksson, W. Li, A. Johlander, et~al. (2016).
Electron jet of asymmetric reconnection.
\textit{Geophys.Res.Lett.}
\textit{43}, 5571–-5580, doi:10.1002/2016GL069064.

\bibitem[\textit{Krall and Liewer}(1971)]{krall71}
Krall, N.A., and P.C. Liewer (1971).
Low-frequency instabilities in magnetic pulses.
\textit{Phys.Rev.A}
\textit{4}, 2094--20103.

\bibitem[\textit{Le Contel et~al.}(2016a)]{lecontel16a}
Le Contel, O., P. Leroy, A. Roux, C. Coillot, D. Alison, A. Bouabdellah, et al. (2016a).
The Search-Coil Magnetometer for MMS.
\textit{Space Sci. Rev.}
\textit{199}, 257--282, doi: 10.1007/s11214-014-0096-9.

\bibitem[\textit{Le Contel et~al.}(2016b)]{lecontel16b}
Le Contel, O., Retin\`{o}, A., H. Breuillard, L. Mirioni, P. Robert, A. Chasapis, et al. (2016b).
Whistler mode waves and Hall fields detected by MMS during a dayside magnetopause crossing.
\textit{Geophys.Res.Lett.}
\textit{43}, 5943--5952, doi:10.1002/2016GL068968.

\bibitem[\textit{Lindqvist et~al.}(2016)]{lindkvist16}
Lindqvist, P.A., G. Olsson, R.B. Torbert, B. King, M. Granoff, D. Rau, et al. (2016).
The Spin-Plane Double Probe Electric Field Instrument for MMS.
\textit{Space Sci. Rev.}
\textit{199}, 137--165, doi: 10.1007/s11214-014-0116-9.

\bibitem[\textit{Norgren et~al.}(2012)]{norgren12}
Norgren, C., A. Vaivads, Yu.V. Khotyainstev, and M. Andr\'e, (2012).
Lower hybrid drift waves: space observations.
\textit{Phys.Rev.Lett.}
\textit{109}, 055001-1--055001-4, doi:10.1103/PhysRevLett.109.055001.

\bibitem[\textit{{\O}ieroset et~al.}(2017)]{oieroset17}
{\O}ieroset, M., T.~D. Phan, M.~A. Shay, C.~C.Haggerty, M. Fujimoto, V. Angelopoulos, et~al. (2017)
THEMIS multispacecraft observations of a reconnecting magnetosheath current sheet with symmetric boundary conditions and a large guide field.
\textit{Geophys.Res.Lett.}
\textit{44}, 7598--7606, doi:10.1002/2017GL074196.

\bibitem[{\textit{Pan et al.}}(2018)]{pan18}
Pan, D-X., Yu.~V. Khotyaintsev, D.~B. Graham, A. Vaivads, X.Z. Zhou, M. Andr\'{e}, et~al. (2018).
Rippled electron-scale structure of a dipolarization front.
\textit{Geophys. Res. Lett.}
\textit{45}, 12116--12124, https://doi.org/10.1029/2018GL080826.

\bibitem[{\textit{Phan et al.}}(2007)]{phan07}
Phan, T.D., G. Paschmann, C. Twitty, F.S. Mozer, J.T. Gosling, J.P. Eastwood, M. {\O}ieroset, H. R\`{e}me, and E.A. Lucek (2007).
Evidence for magnetic reconnection initiated in the magnetosheath.
\textit{Geophys. Res. Lett.}
\textit{34}, L14104, doi:10.1029/2007GL030343.

\bibitem[\textit{Phan et~al.}(2018)]{phan18}
Phan, T.~D., J.~P. Eastwood, M.~A. Shay, J.~F. Drake, B.~U.~\"{O}. Sonnerup, M. Fujimoto, et~al. (2018).
Electron magnetic reconnection without ion coupling in Earth's turbulent magnetosheath.
\textit{Nature}
\textit{57}, 202–-206, https://doi.org/10.1038/s41586-018-0091-5.

\bibitem[\textit{Pollock et~al.}(2016)]{pollock16}
Pollock, C., T. Moore, A. Jacques, J. Burch, U. Gliese,  Y. Saito, et~al. (2016)
Fast Plasma Investigation for Magnetospheric Multiscale.
\textit{Space Sci. Rev.}
\textit{199}, 331--406, doi:10.1007/s11214-016-0245-4.

\bibitem[\textit{Retin\`{o} et~al.}(2007)]{retino07}
Retin\`{o}, A., D. Sundkvist, A. Vaivads, F. Mozer, M. Andr\'e, and C.~J. Owen (2007).
In situ evidence of magnetic reconnection in turbulent plasma.
\textit{Nature Physics}
\textit{3}, 236--238, doi:10.1038/nphys574.

\bibitem[\textit{Russell et~al.}(2016)]{russell16}
Russell, C.~T.,  B.~J. Anderson,  W. Baumjohann, K.~R. Bromund, D. Dearborn, D. Fischer,  et~al. (2016).
The Magnetospheric Multiscale Magnetometers.
\textit{Space Sci. Rev.}
\textit{199}, 189--256, doi: 10.1007/s11214-014-0057-3.

\bibitem[\textit{Santol\'ik et~al.}(2003)]{santolik03}
Santol\'ik, O., M. Parrot, and f. Lefeuvre (2007).
Singular value decomposition methods for wave propagation analysis.
\textit{Radio Sci.}
\textit{38}, 1--13, doi:10.1029/2000RS002523.

\bibitem[\textit{Tang et~al.}(2013)]{tang13}
Tang, X., C. Catell, J. Dombeck, L. Dai, L.~B. Wilson III, A. Breneman, and A. Hupach (2013).
THEMIS observations of the magnetopause electron diffusion region: large amplitude waves and heated electrons.
\textit{Geophys.Res.Lett.}
\textit{40}, 2884--2890, doi: 10.1002/grl.50565.

\bibitem[\textit{Torbert et~al.}(2016)]{torbert16}
Torbert, R.~B., C.~T. Russell, W. Magnes, R.~E. Ergun, P.-A. Lindqvist, O. LeContel,  et~al. (2016).
The FIELDS Instrument Suite on MMS: Scientific Objectives, Measurements, and Data Products.
\textit{Space Sci. Rev.}
\textit{199}, 105--135, doi: 10.1007/s11214-014-0109-8.

\bibitem[{\textit{Torbert et~al.}}(2018)]{torbert18}
Torbert, R.~B., J.~L. Burch, T.~D. Phan, M. Hesse, M.~R. Argall, J. Shuster, J.~P. Eastwood, D.~J. Gershman, P.~A., et~al. (2018).
Electron-scale dynamics of the diffusion region during symmetric magnetic reconnection in space.
\textit{Science}
\textit{362}, 1391--1395, doi: 10.1126/science.aat2998

\bibitem[{\textit{Vaivads et~al.}}(2006)]{vaivads06}
Vaivads, A., Yu. Khotyaintsev, M. Andr\'{e}, and R.~A. Treumann (2006).
Plasma waves near reconnection sites.
\textit{Lect.Notes.Phys.}
\textit{687}, 251--269, doi: 10.1007/3-540-33203-0\_10

\bibitem[\textit{V\"{o}r\"{o}s et~al.}(2016)]{voros16}
V\"{o}r\"{o}s, Z., E. Yordanova, M.M. Echim, G. Consolini, and Y. Narita  (2016).
Turbulence-generated proton-scale structures in the terrestrial magnetosheath.
\textit{Astrophys J Lett}
\textit{819}, L15, doi:10.3847/2041-8205/819/1/L15.

\bibitem[\textit{V\"{o}r\"{o}s et~al.}(2017)]{voros17}
V\"{o}r\"{o}s, Z., E. Yordanova,  A. Varsani, K.~J. Genestreti, Yu.V. Khotyainstev, W.Li, et al.  (2017).
MMS observation of magnetic reconnection in the turbulent magnetosheath.
\textit{J.Geophys.Res. }
\textit{122}, 11442 -- 11467, https://doi.org/10.1002/2017JA024535.

\bibitem[{\textit{Wang et~al.}}(2012)]{wang12}
Wang, C.~P., M. Gkioulidou, L.~R. Lyons, and V. Angelopoulos (2014).
Spatial distributions of the ion to electron temperature ratio in the magnetosheath and plasma sheet.
\textit{J.Geophys.Res.}
\textit{117}, A08215, doi: 10.1029/2012JA017658

\bibitem[{\textit{Wilder et~al.}}(2017)]{wilder17}
Wilder, F.~D., R.~E. Ergun, S. Eriksson, T.~D. Phan, J.~L. Burch, N. Ahmadi, et~al. (2017).
Multipoint measurements of the electron jet of symmetric magnetic reconnection with a moderate guide field.
\textit{Phys.Rev.Lett.}
\textit{118}, 265101--269, doi: 10.1103/PhysRevLett.118.265101

\bibitem[\textit{Yordanova et~al.}(2016)]{yordanova16}
Yordanova, E., Z. V\"{o}r\"{o}s, A. Varsani, D.B. Graham, C. Norgren, Yu.V. Khotyainstev, et al.  (2016).
Electron scale structures and magnetic reconnection signatures in the turbulent magnetosheath.
\textit{Geophys Res Lett }
\textit{43}, 5969 -- 5978, doi:10.1002/2016GL069191.

\bibitem[{\textit{Zhang et~al.}}(1999)]{zhang99}
Zhang, Y., H. Matsumoto, and H. Kojima (1999).
Whistler mode waves in the magnetotail.
\textit{J.Geophys.Res.}
\textit{104} A12, 28633--28644.

\bibitem[{\textit{Zhou et~al.}}(2014)]{zhou14}
Zhou, M., H. Li, X. Deng, S. Huang, Y. Pang, Z. Yuan, et~al. (2014).
Characteristic distribution and possible roles of waves around the lower hybrid frequency in the magnetotail reconnection region.
\textit{J.Geophys.Res.}
\textit{119}, 8228--8242, doi:10.1002/2014JA019978.

\end{thebibliography}
%




\end{document}